\documentclass[twocolumn,epjc3]{svjour3}
\usepackage{epsfig}

\def\ba{\begin{eqnarray}}
\def\ea{\end{eqnarray}}
\def\bq{\begin{equation}}
\def\eq{\end{equation}}
\def\dd{{\mathrm d}}
\def\DD{{\mathcal D}}

\def\MSbar{$\overline{\mathrm{MS}}\ $}

\def\order#1{{\mathcal O}\left(#1\right)}

\def\gsim{\mathrel{\raise.3ex\hbox{$>$\kern-.75em\lower1ex\hbox{$\sim$}}}}
\def\lsim{\mathrel{\raise.3ex\hbox{$<$\kern-.75em\lower1ex\hbox{$\sim$}}}}
\journalname{Eur. Phys. J. C}

\begin{document}

\title{On higher order radiative corrections to elastic electron-proton scattering}
\titlerunning{On higher order rad. corr. to $ep$ scattering}

\author{A.B. Arbuzov\thanksref{e1,addr1,addr2} 
\and T.V. Kopylova\thanksref{addr2} 
}

\thankstext{e1}{e-mail: arbuzov@theor.jinr.ru}

\institute{ \label{addr1} Bogoliubov Laboratory of Theoretical Physics, 
JINR, Dubna,  141980  Russia
\and
\label{addr2} Department of Higher Mathematics, Dubna State University, 
Dubna, 141980  Russia
}

\date{Received: date / Revised version: date}

\maketitle

\begin{abstract}
QED radiative corrections to elastic electron-proton scattering at low energies
are discussed. Corrections to the electron line and effects due to vacuum polarization
are computed. 
Higher order effects are estimated for the conditions of the experiment on
the electric and magnetic proton form factors by A1 Collaboration. Calculations are
performed within the next-to-leading approximation.
Inclusion of the higher order effects can affect the value of the proton charge 
radius extracted from the experimental data.
\end{abstract}


\section{Introduction}

First of all, our paper is motivated by recent very accurate experimental
measurements of the electron-proton elastic scattering at the Mainz Microtron 
(MAMI) \cite{Bernauer:2013tpr}. The average point-to-point errors in the 
cross sections measurement was of the order of a few permille. 

Besides extraction of the proton electromagnetic form factors, the experiment
managed to define the value of the proton electric charge radius with high precision.
It is worth to note that the result for the charge radius extracted from the 
electron-proton scattering data was found to be inconsistent with the one
obtained from muonic hydrogen~\cite{Antognini:1900ns}. The disagreement 
stimulates theoretical studies aimed at its resolution. In the present paper 
we are going to discuss several effects which can affect the data analysis
of low-energy elastic electron-proton scattering. 

The high precision of the experimental measurement of the differential
cross section provides the clear requirement on the magnitude of effects which
should be taken into account. We assume that aiming at the one-permille accuracy,
we have to treat systematically all relative corrections being at least of $10^{-4}$ 
size.

\section{Preliminaries and Notation}

Let us consider the process
\ba \label{proc}
e(p_1) + p(P_1) \longrightarrow e(p_2) + p(P_2) + (n\gamma,\ e^+e^-).
\ea
The initial electron energy $E_1=p_1^0\equiv E$ is of the order 1~GeV, $E\gg m_e$.
The momentum transfer squared $Q^2=-(p_2-p_1)^2$  will be taken in
the range $0.003 < Q^2 < 1$~GeV$^2$ which was explored in the experiment.
Note that the condition $Q^2 \gg m_e^2$ holds for the whole range.
The differential cross section ${d\sigma}/{d\Omega_e}$ of process~(\ref{proc}) 
will be considered.

One-loop QED corrections to the process under consideration are well known.
They are naturally separated into the following parts:\\
--- real and virtual corrections to the electron line, \\
--- real and virtual corrections to the proton line, \\
--- interference of amplitudes of the first two types, \\
--- the effect due to vacuum polarization.
The corresponding analytic results were reproduced in~\cite{Bernauer:2013tpr}
\footnote{Eq.~(14) in this paper contains an obvious misprint: 
the logarithm sign ``$\ln$'' should appear in the first term in square brackets.}
Among one-loop corrections, there is still an open discussion about 
the proper treatment of double photon exchange contributions, see e.g. 
papers~\cite{Arrington:2011dn,Lee:2015jqa} and references therein. We agree with
the importance of this point, but it goes beyond the scope of our 
present study.

To estimate the numerical effect of radiative corrections one has to
take into account concrete experimental conditions. 
Of course, to get the final answer one should include the corrections into
the whole program of the data analysis. 
But our task here will be just to present analytic results with simple
estimates of their impact. So we will simplify the set-up (still following
the main features of the experiment): \\
--- we assume that the measurement is based on the detection of the 
final electron energy and momentum, \\
--- the electron is detected ``bare'', i.e. without possible accompanying photons, \\
--- there is just a simple cut on the lost energy:  $p_1^0-p_2^0\geq \Delta E$
where $\Delta$ is a dimensionless parameter, $\Delta \ll 1$ and $\Delta E \gg m_e$.

The typical magnitude of the $\order{\alpha}$ corrections to the differential
cross section is defined by three major factors: 
\ba
\delta^{(1)} = \frac{\dd\sigma^{(1)}}{\dd\sigma^{(0)}} \sim \frac{\alpha}{2\pi}\cdot 
\ln\left(\frac{Q^2}{m_e^2}\right) \cdot \ln\Delta.
\ea
The enhancement by the so called large logarithm $L\equiv \ln\left(Q^2/m_e^2\right)$
and by the logarithm of the cut-off parameter make the size of the one-loop 
correction to be of the order of a few percent. Since the experimental uncertainties 
are well below this order, the one-loop corrections were treated in the
data analysis with care, see details in~\cite{Bernauer:2013tpr}.

The purpose of our paper is to estimate the leading and next-to-leading
higher order corrections. We will consider one by one the following 
higher order contributions: \\
1. higher order effects in vacuum polarization; \\
2. cut-off dependence of the photonic corrections; \\
3. light pair corrections in the leading logarithmic approximation; \\
4. complete next-to-leading  $\order{\alpha^2L^1}$ corrections to the lepton line.

As can be seen from the first order, higher order corrections only to the electron line
and to vacuum polarization can be numerically important.

\subsection{Higher order effects in vacuum polarization}

Running of the QED coupling constant can be naturally represented as
\begin{eqnarray} \label{eq:Da}
&&  \alpha(Q^2) = \frac{\alpha(0)}{1-\Pi(Q^2)}, 
\\ \nonumber
&& \Pi(Q^2) = \Pi_e(Q^2) + \Pi_\mu(Q^2) + \Pi_{\mathrm{hadr}}(Q^2) + \dots
\end{eqnarray}
where $\alpha(0)\equiv\alpha\approx 1/137.036$.
A discussion of the relative size of different contributions to $\Pi(Q^2)$ for low $Q^2$ 
values can be found in Ref.~\cite{Arbuzov:2004wp}. The magnitude of $\Pi(Q^2)$ for the 
range of momentum transfer under consideration is about $0.01$.
The bulk of the vacuum polarization effect comes
from one-loop $e^+e^-$ pair insertion into the photon propagator,
\ba
&& \Pi_{e}(Q^2) = \frac{\alpha(0)}{\pi} \biggl(\frac{1}{3}L-\frac{5}{9}\biggr)
+\biggl(\frac{\alpha(0)}{\pi}\biggr)^2\biggl(\frac{1}{4}L
\nonumber \\ && \quad 
+ \zeta(3) - \frac{5}{24} \biggr) + {\mathcal{O}}(\alpha^3).
\ea
One can note that the $\order{\alpha^2}$ contribution is of the next-to-leading order, 
since it contains only the first power of the large logarithm $L$. So it makes only
a $\sim 10^{-5}$ effect well below the precision tag.
The resummation of the vacuum polarization effect gives
\ba \label{geom}
\delta\sigma_{\mathrm{vac.pol.}}=\sigma^{(0)}\left(\frac{\alpha(Q^2)}{\alpha(0)}\right)^2
= \frac{\sigma^{(0)}}{|1-\Pi(Q^2)|^2}.
\ea


Polarization of vacuum by virtual $\mu^+\mu^-$ pairs is not as large 
as by the $e^+e^-$ ones. But in the bulk of the kinematical domain
the suppression is only logarithmic. So,
\ba
&& \Pi_\mu(Q^2) = \frac{\alpha}{\pi}\biggl[
\frac{v_\mu}{2}\left(1-\frac{v_\mu^2}{3}\right)\ln\frac{v_\mu+1}{v_\mu-1}
+ \frac{v_\mu^2}{3} - \frac{8}{9}\biggr]
\nonumber \\ && \quad
+\order{\alpha^2}, \quad
v_\mu \equiv \sqrt{1+\frac{4m_\mu^2}{Q^2}}
\ea
has to be taken into account at least in the first order in $\alpha$.
For $Q^2=1$~GeV it reaches $2\cdot 10^{-3}$.

Instead of the resummed geometrical series of Eq.~\ref{geom}, 
the A1 collaboration in Ref.~\cite{Bernauer:2013tpr} 
used exponentiation of the effect of the vacuum polarization by leptons, which 
is close numerically for the given $Q^2$ range, see Fig.~\ref{vac_diff} below. 

The hadronic contribution $\Pi_{\mathrm{hadr}}(Q^2)$ 
is rather small at $Q^2 \leq 1$~GeV, but at the right edge
it is rising steeply and reaches a few permille. 
Contributions of tau leptons and electroweak bosons are obviously 
numerically negligible in our case.
More detailed numerical estimates of vacuum polarization effects 
will be presented below in Sect.~\ref{NumRes}.

\subsection{Cut-off dependence of the photonic corrections}

The Yennie-Frautschi-Suura theorem~\cite{Yennie:1961ad} proves that emission
of each soft photon can be treated as an independent process. As the result,
multiple emission of soft photons can be resummed into an exponent. 
By construction in the case of independent emission of soft photons,
the maximal energy of each photon is limited independently.
But in the given experimental set-up, we have a cut-off on the total lost
energy. The corresponding effect was considered {\it e.g.} in Ref.~\cite{Arbuzov:1998du}.
For double soft photon emission in gives the following shift:
\ba
\mathrm{e}^{\delta_{\mathrm{soft}}} \to \mathrm{e}^{\delta_{\mathrm{soft}}}
- \left(\frac{\alpha}{\pi}\right)^2\frac{\pi^2}{3}\left(L-1\right)^2.
\ea
At $Q^2=1$~GeV$^2$ this leads to a visible relative shift of the cross section
of about $-3.5\cdot 10^{-3}$.

To have the theoretical precision under control we can estimate
the effect also for the leading logarithmic photonic correction 
in the third order.
The relative correction reads
\ba \label{d2sig3}
&& \delta^{(3)}_{\mathrm{LLA}}= (L-1)^3 \biggl(\frac{\alpha}{\pi}\biggr)^3 
\frac{1}{6}\left(P^{(0)}\otimes P^{(0)}\otimes P^{(0)}\right)_{\Delta},
\nonumber \\ && 
\left(P^{(0)}\otimes P^{(0)}\otimes P^{(0)}\right)_{\Delta}
= 8\left(P^{(0)}_{\Delta}\right)^3 - 24\zeta(2)P^{(0)}_{\Delta} 
\nonumber \\ && \quad
+ 16\zeta(3).
\ea
So, the treatment of the cut-off results in the relative shift of the order
\ba
\delta^{(3)}_{\mathrm{cut}} = (L-1)^3 \biggl(\frac{\alpha}{\pi}\biggr)^3
\biggl[ - 4\zeta(2)P^{(0)}_{\Delta} + \frac{8}{3}\zeta(3) \biggr]
\ea
which is not small and reaches about $2\cdot 10^{-3}$.

In the same way one can verify that the na\"ive exponentiation leads to
a considerable off-set in the fourth order leading logarithmic correction:
\ba
\delta^{(4)}_{\mathrm{cut}} &=& (L-1)^4 \biggl(\frac{\alpha}{\pi}\biggr)^4
\biggl[ - 24\zeta(2)\left(P^{(0)}_{\Delta}\right)^2 
\nonumber 
+ \frac{4\pi^4}{15}\biggr]
\ea
which again is of the order of a few times $10^{-3}$. Meanwhile, the total 
effect of the fourth order leading log correction in the considered
kinematical domain does not exceed $1\cdot10^{-4}$. The explicit expression for
convolution of four splitting functions, which appear in
\ba
\delta^{(4)}_{\mathrm{LLA}} = (L-1)^4 \biggl(\frac{\alpha}{\pi}\biggr)^4
\frac{1}{24}\left(P^{(0)}\right)^{\otimes 4}_{\Delta},
\ea 
can be found in Ref.~\cite{Arbuzov:1999cq}.

The proper exponentiation of radiative corrections in the leading logarithmic
approximation is based on the exact solution of the renormalization group equation,
see~\cite{Kuraev:1985hb}. But for the practical application under consideration it
is sufficient to compute effect order by order and keep the theoretical uncertainty
under control in this way.

\subsection{Light pair corrections the leading logarithmic approximation}

The contribution of $e^+e^-$ pairs can be easily estimated with
the help of the leading logarithmic approximation (LLA) 
in QED~\cite{Kuraev:1985hb,Skrzypek:1992vk,Arbuzov:2001rt}:
\ba \label{pair_LLA}
& \delta_{\mathrm{pair}}^{LLA} = \frac{2}{3}\left(\frac{\alpha}{2\pi}L\right)^2P^{(0)}_\Delta
+ \frac{4}{3}\left(\frac{\alpha}{2\pi}L\right)^3
\biggl\{\left(P^{(0)}\otimes P^{(0)}\right)_{\Delta} 
\nonumber \\ & \quad 
+ \frac{2}{9}P^{(0)}_\Delta \biggr\}
+ \order{\alpha^2L,\alpha^4L^4}
\ea
where the so-called $\Delta$-parts of 
splitting functions (see e.g. Refs.~\cite{Skrzypek:1992vk,Arbuzov:1999cq})
read
\ba
&& P^{(0)}_{\Delta} = 2 \ln\Delta +\frac{3}{2}, 
\nonumber \\
&& \left(P^{(0)}\otimes P^{(0)}\right)_{\Delta}
= \left( P^{(0)}_{\Delta}\right)^2 - \frac{\pi^2}{3}.
\ea
Note that in the third order in $\alpha$ we have an effect due to
simultaneous (either virtual or soft) radiation of a pair and a photon.

To have a better control on the precision level, we can include also
the next-to-leading pair corrections in the order $\order{\alpha^2L}$
where some enhancement due to the experimental cut-off takes place.
The corresponding effect will be estimated below.

\subsection{Complete next-to-leading logarithmic corrections to the lepton line}

In order to control the precision of theoretical estimates
we can compute the complete set of next-to-leading order (NLO)
corrections to the given process by means of the renormalization
group approach to QED~\cite{Kuraev:1985hb}. 
The NLO QED structure functions were first introduced
in~\cite{Berends:1987ab}. The corresponding fragmentation functions
were used in~\cite{Arbuzov:2002cn,Arbuzov:2002rp} to evaluate NLO
corrections to the muon decay spectrum. Here we can follow the
paper~\cite{Arbuzov:2006mu}, where NLO QED corrections were
computed in a similar set-up for the case of Bhabha scattering.

The relevant photonic and $e^+e^-$ pair contributions to the NLO electron structure (str) 
and fragmentation (frg) functions have the form\footnote{We dropped the
singlet channel contributions which are suppressed in the given experimental set-up.}
\ba  \label{Dee}
&& \DD_{ee}^{\mathrm{str,frg}} (z) = \delta(1-z)
+ \frac{\alpha}{2\pi}\biggl( LP^{(0)}(z)
+ d_1(z) \biggr)
 \nonumber \\ && \quad
+ \biggl(\frac{\alpha}{2\pi}\biggr)^2\biggl(
\frac{1}{2}L^2P^{(0)}\otimes P^{(0)}(z)
+\frac{1}{3}L^2P^{(0)}(z)
\nonumber \\ && \quad
+ LP^{(0)}\otimes d_1(z)
+ LP^{(1,\gamma){\mathrm{str,frg}}}_{ee}(z) 
\nonumber \\ && \quad
+ LP^{(1,\mathrm{pair}){\mathrm{str,frg}}}_{ee}(z) \biggr)
+ \order{\alpha^2 L^{0}, \alpha^3}.
\ea
Explicit expressions for splitting functions $P^{(n)}_{ee}$
and $d_1$ can be found in~\cite{Arbuzov:2006mu}. The master
formula for NLO photonic corrections to elastic
electron-proton scattering reads
\ba \label{master}
&& \dd \sigma = \int^{1}_{\bar{z}} \dd z \DD^{\mathrm{str}}_{ee} (z) 
 \biggl( \dd \sigma^{(0)} (z) + \dd \bar{\sigma}^{(1)} (z) 
\nonumber \\ && \quad
+ \order{\alpha^2L^0}  \biggr) 
\int^{1}_{\bar{y}} \frac{\dd y}{Y} 
\DD^{\mathrm{frg}}_{ee} \left(\frac{y}{Y}\right), 
\ea
where $\dd\bar\sigma^{(1)}$ is the $\order{\alpha}$ correction to the $ep$ scattering with
a ``massless electron'',
calculated using the \MSbar scheme to subtract the lepton mass singularities.  
The energy fraction of the incoming parton is $z$, and 
$Y$ is the the energy fraction of the outgoing (observed) electron.
As concerning the factorization scale, it is natural to choose it to
be equal to the momentum transfer: $L\equiv \ln(Q^2/m_e^2)$.

Here we are interested in the contributions due to virtual and soft photons, so 
both integrals have the same lower limit being equal to $1-\Delta$.
First we can perform convolution of the structure and fragmentation functions entering
Eq.~(\ref{master}) with each other 
$\DD^{\mathrm{str}}_{ee}\otimes \DD^{\mathrm{frg}}_{ee} (z)$. 
If $z=1-\Delta$ and $\Delta\ll 1$, the result of the convolution gives the probability 
density to find such a situation where one looses in total due to photon emission 
$\Delta E_{\mathrm{beam}}$ from the total energy of the process under consideration.

Convolution of the function found above with the Born part of the kernel cross section 
gives us the corresponding part
to the cross section (with the upper limit on the lost energy):
\ba \label{d2sig0}
&&\dd\sigma^{\mathrm{NLO}}= \int^1_{1 - \Delta} \DD^{\mathrm{str}}_{ee}\otimes \DD^{\mathrm{frg}}_{ee}(z) 
\nonumber \\ && \quad
\times \biggl[ \dd \sigma^{(0)}(z) 
+ \dd \bar{\sigma}^{(1)}(z) \biggr] \dd z 
\nonumber \\ && \quad
=  \dd\sigma^{(0)}(1) \Biggl\{ 1 + 2\frac{\alpha}{2 \pi} 
\biggl[  LP^{(0)}_{\Delta} + (d_1)_{\Delta}  \biggr]  
\nonumber \\ && \quad 
+ 2\biggl(\frac{\alpha}{2 \pi}\biggr)^2 \biggl[ 
  L^2 \left(P^{(0)}\otimes P^{(0)}\right)_{\Delta}
+ \frac{1}{3}L^2 P^{(0)}_{\Delta}
\nonumber \\ && \quad 
+ 2 L (P^{(0)} \otimes d_1)_{\Delta} 
+ L (P^{(1,\gamma)}_{ee})_{\Delta}  
+ L (P^{(1,\mathrm{pair})}_{ee})_{\Delta}  
\biggr] \Biggr\}
\nonumber \\ && \quad 
+  \dd \bar{\sigma}^{(1)}(1)\, 2\frac{\alpha}{2\pi}
LP^{(0)}_{\Delta}  + \order{\alpha^3L^3}
\ea
where the relevant $\Delta$-parts read
\ba 
&& (d_1)_{\Delta} = - 2\ln^2\Delta - 2\ln\Delta + 2,
\nonumber \\
&& (P^{(0)} \otimes d_1)_{\Delta} =  - 4\ln^3\Delta
-7\ln^2\Delta + \ln\Delta\left(1+8\zeta(2)\right)
\nonumber \\ && \quad
+ 3 - 8\zeta(3) + 4\zeta(2),
\nonumber \\
&& (P^{(1,\gamma)}_{ee})_{\Delta} = \frac{3}{8} - 3\zeta(2) + 6\zeta(3),
\nonumber \\
&& (P^{(1,\mathrm{pair})}_{ee})_{\Delta} = - \frac{20}{9}\ln\Delta 
- \frac{1}{6}  - \frac{4}{3}\zeta(2).
\ea
The values of the Riemann zeta function are
$\zeta(2)=\pi^2/6$ and $\zeta(3)\approx 1.202$.
The $\Delta$-parts of the structure and fragmentation
splitting functions $(P^{(1,\gamma(\mathrm{pair})){\mathrm{str}}}_{ee})_{\Delta}$ and 
$(P^{(1,\gamma(\mathrm{pair})){\mathrm{frg}}}_{ee})_{\Delta}$
coincide, so the notation is simplified.

Note that by construction in the \MSbar scheme, the complete first order correction 
is reproduced since
\ba
\dd\bar{\sigma}^{(1)}(1) &=& \dd\sigma^{(1)}(1) 
\nonumber \\ 
&-& 2\dd\sigma^{(0)}(1)\frac{\alpha}{2\pi}\biggl[ LP^{(0)}_\Delta
+ (d_1)_\Delta \biggr].
\ea
The factor 2 before the subtracted term on the right hand side 
reflects the presence of mass singularities in both the initial and final
state corrections.

\section{Numerical results \label{NumRes}}

\begin{figure}
\includegraphics[width=7.5cm,height=5cm,angle=0]{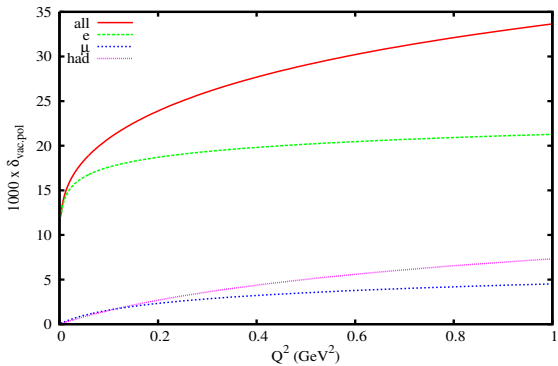}
\caption{Vacuum polarization corrections due to electrons ($e$), muons ($\mu$), 
hadrons (had), and the combined effect (all).
\label{delta_vp}}
\end{figure}

Fig.~\ref{delta_vp} shows different contributions to the vacuum polarization correction
\ba
\delta_{\mathrm{vac.pol.}} = \frac{\delta\sigma_{\mathrm{vac.pol.}}}{\sigma^{(0)}}.
\ea
This figure was obtained with the help of the Fortran package {\tt alphaQED} 
by F.~Jegerlehner~\cite{Jegerlehner:2011mw}.
One can see that vacuum polarization by muons and hadrons contributes by up to one percent.
That is a rather large effect for the given precision tag. Moreover, the momentum dependence
of the total vacuum polarization correction is different from the pure electron one.
That can affect the extrapolation procedure which is applied for extraction of the proton
charge radius.

As concerning the hadronic contribution to vacuum polarization it can be either treated
as a part of radiative corrections or as a part of the proton form factor. To our mind,
the former treatment has two advantages. First, this contribution is always there as for
point-like as well as for non-point-like particles. Second, in higher order corrections
it is not factorized out as can be seen already in Eq.~(\ref{geom}). 
From the first glance the hadronic contribution should not affect the value of the proton
charge radius since it is defined at the zero momentum transfer, where this effect is vanishing.
Nevertheless, the effect has a pronounced $Q^2$ dependence in the explored domain and
it certainly affects the extrapolation to the zero momentum transfer point. For this reason
we recommend to treat the hadronic vacuum polarization as a part of radiative corrections
along this the corresponding leptonic contributions.

\begin{figure}
\includegraphics[width=7.5cm,height=5cm,angle=0]{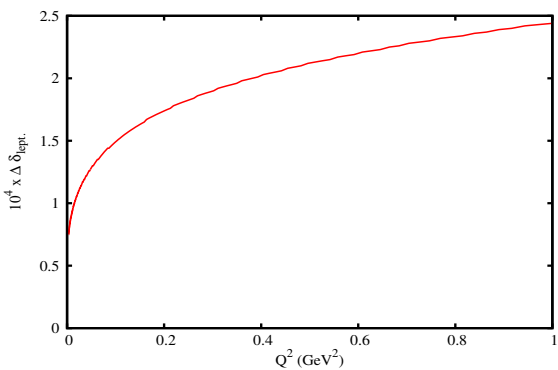}
\caption{Difference in the leptonic vacuum polarization corrections.
\label{vac_diff}}
\end{figure}

Fig.~\ref{vac_diff} shows the difference between the corrections due to vacuum 
polarization by electron and muons between the result obtained with the
help of the Fortran package {\tt alphaQED} (taking into account also known 
2-loop contributions) and the exponentiated treatment of the effect described 
in Ref.~\cite{Bernauer:2013tpr}. One can see that the difference is of the order
of $2\cdot 10^{4}$ which might be relevant for a better control of systematic 
errors.

\begin{figure}
\includegraphics[width=7.5cm,height=5cm,angle=0]{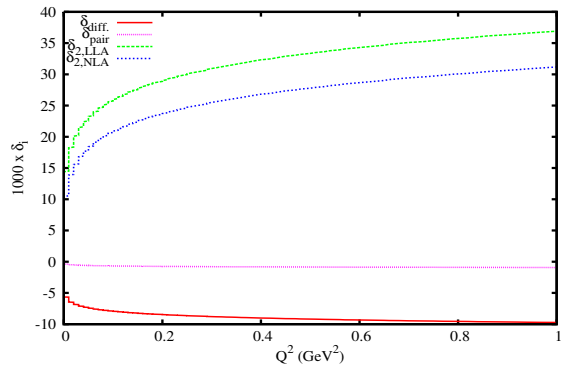}
\caption{Relative higher order QED corrections to electron line in $ep$ scattering cross section {\it vs} momentum transferred squared.
\label{delta_ho}}
\end{figure}

Relative QED corrections to the electron line
\ba
\delta_i = \frac{\dd\sigma^{(i)}}{\dd\sigma^{(0)}}
\ea
are presented in Fig.~\ref{delta_ho}. Index $i$ runs over: \\
a) ``2,LLA'', i.e. pure photonic $\order{\alpha^2L^2}$ corrections from Eq.~(\ref{d2sig0}), \\
b) ``2,NLA'', i.e. the sum of pure photonic $\order{\alpha^2L^2}$ and $\order{\alpha^2L^1}$ 
corrections from Eq.~(\ref{d2sig0}), \\
c) ``pair'', i.e. the leading log pair corrections from Eq.~(\ref{pair_LLA}) 
supplemented by subleading pair corrections extracted from Eq.~(\ref{d2sig0}), \\
d) ``diff.'', i.e. the shift from the exponentiated one-loop result:
\ba \label{d_diff}
\delta_{\mathrm{diff.}} &=& \frac{\dd\sigma^{\mathrm{NLO}}}{\dd\sigma^{(0)}(1)}   
+ \delta^{(3)}_{\mathrm{LLA}} 
+ \delta^{(3)}_{\mathrm{LLA,pair}} 
+ \delta^{(4)}_{\mathrm{LLA}} 
\nonumber \\ 
&-& \exp\{\delta^{(1)}\}.
\ea

\section{Conclusions}

In this way we presented results for higher order corrections to 
elastic electron-proton scattering which can be relevant for modern
high-accuracy experiments. The corrections are presented in an
analytic form. Numerical results are given for a simplified experimental
set-up just to estimate the magnitude of effects. 
Matching with exponentiated representation of corrections 
is straightforward, since we have explicit results for sub-leading corrections.

Quantity~(\ref{d_diff}) plotted in Fig.~\ref{delta_ho} is an estimate of the effect 
due to an advanced treatment of higher order corrections to the electron line
in the process of $ep$ scattering, which is presented here.
We have shown also that accurate treatment of vacuum polarization effects is
also important for getting a high precision.
An adequate treatment of all other relevant effects (double photon exchange,
radiative corrections to the proton line, details of the experimental set-up,
{\it etc.}) is also required.

\begin{acknowledgement}
We are grateful to S.G. Karshenboim for pointing out the problem and useful discussions.
\end{acknowledgement}

\end{document}